\documentclass[aps,prb,twocolumn,floatfix,superscriptaddress]{revtex4-2}
\usepackage{amssymb,graphicx,color}
\usepackage[intlimits]{amsmath}
\usepackage[english]{babel}
\usepackage{xr-hyper} 
\usepackage[colorlinks]{hyperref}
\usepackage{soul}
\hypersetup{
	colorlinks=blue,
	linkcolor=blue,
	filecolor=blue,
	urlcolor=blue,
	citecolor=blue
}
\usepackage[normalem]{ulem}
\usepackage{comment}
\usepackage{ragged2e}
\usepackage{enumerate}
\newcommand{\nc}{\newcommand}
\nc{\nn}{\nonumber}
\nc{\txt}{\textrm}
\nc{\txtsup}{\textsuperscript}
\nc{\txtsub}{\textsubscript}
\nc{\calL}{\mathcal{L}}
\nc{\U}{\mathcal{U}}
\nc{\T}{\mathcal{T}}
\nc{\E}{\mathcal{E}}
\nc{\calH}{\mathcal{H}}
\nc{\calD}{\mathcal{D}}
\nc{\SARKAR}[1]{\textcolor{blue}{#1}}
\nc{\DB}[1]{\textcolor{red}{#1}}
\nc{\BKA}[1]{\textcolor{magenta}{#1}}
\nc{\hatc}{\hat{c}}
\nc{\mb}{\mathbb}
\begin{document}

\title{Impact of dephasing on non-equilibrium steady-state transport in fermionic chains with long-range hopping} 

\author{Subhajit Sarkar}
\email{subhajit.sarkar@uj.edu.pl}
\affiliation{Institute of Theoretical Physics, Jagiellonian University, \L{}ojasiewicza 11, 30-348 Krak\'{o}w, Poland.}

\author{Bijay Kumar Agarwalla}
\email{bijay@iiserpune.ac.in}
\affiliation{Department of Physics, Indian Institute of Science Education and Research Pune, Dr. Homi Bhabha Road, Ward No. 8, NCL Colony, Pashan, Pune, Maharashtra 411008, India.}

\author{Devendra Singh Bhakuni}
\email{devendra@post.bgu.ac.il}
\affiliation{The Abdus Salam International Centre for Theoretical Physics (ICTP), Strada Costiera 11, 34151 Trieste, Italy}
\affiliation{Department of Physics, Ben-Gurion University of the Negev, Beer Sheva 84105, Israel.}

\begin{abstract}
Quantum transport in a non-equilibrium setting plays a fundamental role in understanding the properties of systems ranging from quantum devices to biological systems. Dephasing- a key aspect of out-of-equilibrium systems, arises from the interactions with the noisy environment and can profoundly modify transport features. Here, we investigate the impact of dephasing on the non-equilibrium steady-state transport properties of non-interacting fermions on a one-dimensional lattice with long-range hopping ($\sim \frac{1}{r^\alpha}$). We show the emergence of distinct transport regimes as the long-range hopping parameter $\alpha$ is tuned. In the short-range limit ($\alpha \gg 1$), transport is diffusive, while for the long-range limit ($\alpha \sim \mathcal{O}(1)$), we observe a super-diffusive transport regime. Using the numerical simulation of the Lindblad master equation, and corroborated with the analysis of the current operator norm, we identify a critical long-range hopping parameter, $\alpha_c \approx 1.5$, below which super-diffusive transport becomes evident that quickly becomes independent of the dephasing strength. Interstingly, within the super-diffusive regime, we find a crossover from logarithmic to power-law system-size dependence in the non-equilibrium steady-state resistance when $\alpha$ varies from $\alpha \leq 1$ to $\alpha \lesssim 1.5$. Our results, thus, elucidate the intricate balance between dephasing and unitary dynamics, revealing novel steady-state transport features.
\end{abstract}

\maketitle

\section{Introduction}
Quantum transport of charge and energy in non-equilibrium settings stands as a foundational cornerstone in modern physics. This includes phenomena such as light-harvesting in photosynthesis and other chemical processes in both chemical and biological systems~\cite{Harush_dubi_Sci_Adv, mattioni2021design}. It further extends to the exploration of emerging non-equilibrium states of matter in nano-fabricated quantum devices~\cite{Maier_PhysRevLett_ENAQT_10_qubit, Barthelemy_2013, sarkar2022emergence, sarkar2022signatures}. Specifically, charge transport measurements using quantum dot arrays have led to a multitude of breakthroughs in quantum physics. These encompass phenomena such as Coulomb blockade, Kondo physics, superconductivity, quantum-point-contact universal conductance, and even aspects of quantum computing~\cite{Field_PhysRevLett, Byrnes_PhysRevB2008, Barthelemy_2013, Deshpande2010, Loss_DiVincenzo_PhysRevA}.

The study of quantum transport often focuses on analyzing the behavior of the mean squared displacement of a wave packet or observing the flow of energy and charge through a system using a boundary drive~\cite{steinigeweg_2009,bar_lev_absence_2015,steinigeweg_real-time_2017,luitz_ergodic_2017,znidaric_2010,znidaric_2013,znidaric_2017}. For example, for systems displaying ballistic transport, the mean squared displacement of a wave packet grows quadratic in time, while in diffusive systems, its growth is linear in time. Alternatively, for the boundary-driven systems, if the current in the non-equilibrium steady state (NESS) decreases by the inverse of system size, i.e., $J_{\infty}(L) \sim L^{-1}$, or remains constant as the system size increases, it indicates diffusive or ballistic transport, respectively. The subscript `$\infty$' in $J_{\infty}(L)$ here refers to the current evaluated in the steady state. Any deviation from these scaling of mean squared displacement or steady state current signifies anomalous transport \cite{Bertini_RevModPhys, eisler2011crossover, esposito2005emergence, esposito2005exactly}.

Building on this, dephasing, introduced via coupling to an environment, profoundly impacts the nature of quantum transport in lattice systems \cite{znidaric_2010,znidaric_2013,znidaric_2017,talia2022logarithmic,bhakuni2023noise}. In particular, a Zeno-type dephasing characterized by classical noise~\cite{Gu_JCP_noise_2019} can give rise to a phenomenon known as environment-assisted quantum transport~\cite{Harush_dubi_Sci_Adv, zerah2020effects, rebentrost2009environment, sarkar2020environment}. Additionally, studies employing dynamical measures have demonstrated that dephasing can cause anomalous transport in localized systems, manifesting as sub-diffusive, diffusive, or logarithmic behavior based on the degree of noise sparsity~\cite{gopalakrishnan2017noise, talia2022logarithmic, bhakuni2023noise}. In relation to a boundary-driven system in the nearest-neighbor chain of free fermions, recent investigations point to a crossover from ballistic to diffusive transport by varying dephasing strength~\cite{Turkeshi_PhysRevB, Madhumita}.

While most research has focused on systems with short-range hopping, how transport properties of systems exhibiting hopping beyond nearest-neighbor changes in the presence of dephasing remains an open question. These long-range systems, characterized by interactions/hopping that decay as $r^{-\alpha}$ (where $r$ is the distance between two sites), are widely observed in nature and can also be engineered in cold-atom and trapped ionic systems \cite{richerme2014nonlocal, jurcevic2014quasiparticle, gring2012relaxation, morong2021observation, kyprianidis2021observation}. The presence of long-range hopping leads to qualitative changes in various physical properties, including the equilibrium phase, ground state, and dynamic properties~ \cite{Kosterlitz_PhysRevLett, kuwahara2020area, koffel2012entanglement, vodola2014kitaev, vodola2015long, chen2019finite, tran2020hierarchy, Kuwahara_light_cones2020, Zhou_levy_flight2020, Kuwahara_noscrambling2020, Sondhi_MBL_long_PhysRevX}. For $\alpha>d$, with $d$ being the dimension of the system, the physics often remains qualitatively the same compared to the short-range systems. However, the opposite limit of $\alpha<d$ can lead to exciting and novel features such as logarithmic growth of entanglement \cite{schachenmayer2013entanglement, lerose2020origin}, heating suppression~\cite{Devendra_PhysRevB_long_range}, light-cone evolution \cite{santos2016cooperative, storch2015interplay}, and self-trapping \cite{nazareno1999long}. 

In this work, we investigate the NESS transport properties of the long-range systems coupled to dissipative particle injection-extraction baths with a rate $\Gamma$ corresponding to an infinite bias condition and a dephasing environment with a rate $\gamma$. Note that previous work focused on these systems without dephasing, and a sub-diffusive scaling of conductance was observed under proper tuning of chemical potential bias \cite{Purkayastha_subdiffusive_2021}. A finite bias window also leads to a substantial non-Markovianity \cite{Gabriela_PhysRevA_1, wojtowicz2023accumulative}, a situation that needs further investigation.

Here, we analyze the Markovian Lindblad quantum master equation in the steady state. Remarkably, we observe that above a critical value of the long-range parameter, $\alpha_c \approx 1.5$, the NESS charge current exhibits a diffusive scaling with the system size, i.e., $ J_{\infty} \sim L^{-1}$. However, beneath the critical parameter $\alpha_{c}$, we observe a captivating super-diffusive transport regime. This regime displays a power-law ($J_{\infty} \sim L^{-\nu}$, with $\nu <1$) dependence on system size for $1 <\alpha <1.5$. For $\alpha \leq 1 $, this super-diffusive transport regime displays a logarithmic dependence of the current on the system size ($ J_{\infty} \sim [\log(L)]^{-1}$), overall indicating an anomalous transport. The emergence of distinctive transport regimes can be attributed to the intriguing interplay between long-range hopping and dephasing, a detail we will delve into in subsequent results sections. Building on these findings, we also highlight the striking resemblance between the NESS current heat map in the $\alpha-\gamma$ plane with the entanglement phase diagram of the measurement-induced phase transition (MIPT) recently illustrated in long-range hopping systems~\cite{block2022measurement, muller2022measurement, minato2022fate}.


The organization of the paper is as follows: We introduce the model Hamiltonian and provide a formulation of the characterization of the non-equilibrium transport in Sec.\ref{Sec:model}. We provide the central findings of our work in Sec.~\ref{Sec:results}. Finally, in the last section, we summarize and discuss the possible connection with MIPT.

\section{Model Hamiltonian and Formulation}\label{Sec:model}
\begin{figure}
    \centering
    \includegraphics[scale=0.085]{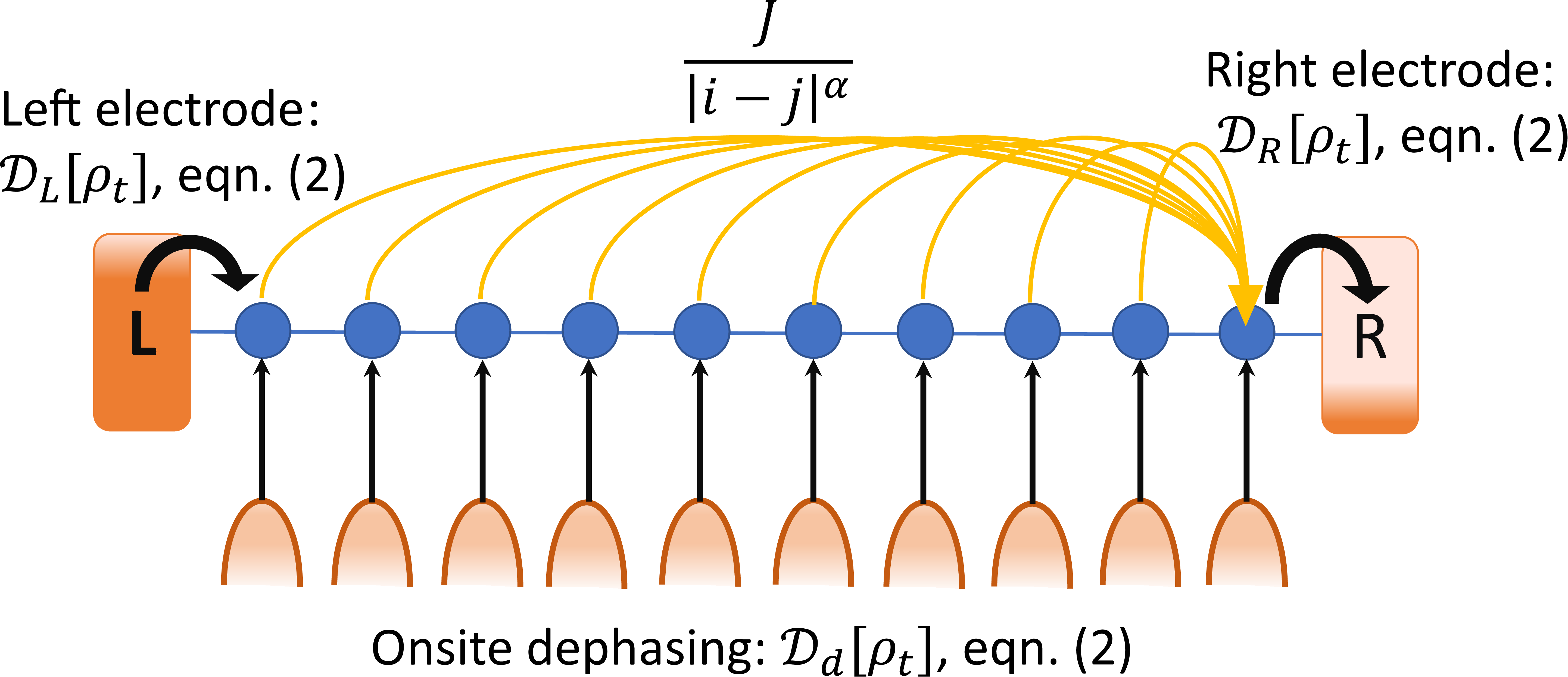}
    \caption{(Color online:) Schematic of our set-up: a non-interacting fermionic lattice chain with long-range hopping {(indicated by yellow solid lines)} is placed between two baths (electrodes kept at the infinite bias limit), `L'-- left and `R'--right injecting and extracting particles at the first and last sites, respectively. At each site of the chain, an on-site dephasing is applied that mimics the Zeno-type measurement of number density at that site, indicated by arrows.}
    \label{fig:schematic}
\end{figure}

We consider a linear chain of $L$ sites hosting non-interacting spinless Fermions with long-range hoping, as shown in Fig.~\ref{fig:schematic}. The model Hamiltonian {for the chain} is given by
\begin{eqnarray}\label{eq:ham}
\hat{\calH} &=& \sum_{j=1}^{L-1} \left[ \sum_{r} \frac{J}{r^\alpha} (\hatc_{j}^{\dagger} \hatc_{j+r} + \text{h.c.})\right]
= \sum_{i,j=1}^{L-1} \sum_{r} \hatc_{i}^{\dagger}  \mb{H}_{i,j}^{(r)} \hatc_{j} \nn \\
& & ~\text{with}~\mb{H}_{i,j}^{(r)} = \frac{J}{r^\alpha}(\delta_{i,j+r}+ \delta_{i+r,j}), 
\end{eqnarray}
where $\frac{J}{r^\alpha}$ is the long-range hopping strength with the exponent of the spatial dependence $\alpha$, $r = |i-j|$ being the distance between lattice sites $i$ and $j$
. 
The operator norm of the long-range hopping term is given by $L^{1-\alpha}$, and for $\alpha < 1$, it becomes dominant in the thermodynamic limit \cite{defenu_long_range_pnas}. However, one can re-scale such a term by an $\alpha-$dependent factor $\mathcal{N}_{\alpha}$ factor which is $\mathcal{N}_{\infty} = 2$ and $\mathcal{N}_{0} = L$ \cite{kastner2011diverging, kastner2017nscaling, bachelard2013universal, Jin_PhysRevResearch_quantum_resistor}, experiments are indeed performed with a finite number of sites where this re-scaling does not appear naturally \cite{Maier_PhysRevLett_ENAQT_10_qubit, richerme2014nonlocal, jurcevic2014quasiparticle, neyenhuis2017observation}. Therefore, we do not consider this re-scaling in our work, following Ref. \cite{Devendra_PhysRevB_long_range}.

We couple the system to a particle injecting source and a particle extraction sink to the left and right end, respectively, with a constant rate $\Gamma$.  Additionally, we add Zeno-type dephasing at each site with strength $\gamma$ providing an energy relaxation channel. The equation of motion of the density matrix of the system is governed by the standard Lindblad quantum master equation.
\begin{eqnarray}\label{eq:eqn_mot}
\frac{\partial \rho_{t}}{\partial t} &=& -i [\hat{\calH}, \rho_{t}] + \calD_{d}[\rho_{t}]+\calD_{L}[\rho_{t}]+\calD_{R}[\rho_{t}] \nonumber \\
\calD_{d}[\rho_{t}] &=& \frac{\gamma}{2} \sum_{j=1}^{L} \left(\hat{n}_{j}\rho_t \hat{n}_{j} - \frac{1}{2}\lbrace  \hat{n}_{j}, \rho_{t} \rbrace \right)\nonumber \\
\calD_{L}[\rho_{t}] &=& \frac{\Gamma}{2}  \left(\hat{c}_{1}^{\dagger}\rho_t \hat{c}_{1} - \frac{1}{2}\lbrace \hat{c}_{1} \hat{c}_{1}^{\dagger}, \rho_{t} \rbrace \right)\nonumber \\
\calD_{R}[\rho_{t}] &=& \frac{\Gamma}{2} \left(\hat{c}_{L} \rho_t \hat{c}_{L}^{\dagger} - \frac{1}{2}\lbrace \hat{c}_{L}^{\dagger}\hat{c}_{L} , \rho_{t} \rbrace \right),
\end{eqnarray}
where $\calD_{d}[\rho_{t}]$, $\calD_{L}[\rho_{t}]$ and $\calD_{R}[\rho_{t}]$ are the Lindblad dissipators correspond to the on-site dephasing, left and right boundary drive, respectively, and $\hat{n}_{i}$ is the Fermion number operator at site $i$.

For the long-range Hamiltonian, the local current is defined from the particle number conservation \cite{sarkar2020environment}: $\displaystyle \frac{d \hat{n}_{j}}{dt} = \hat{J}_{j,in} - \hat{J}_{j,out}$. For the site connected to the left-lead, the current into the first site is, $\hat{J}_{1,in} = \calD_{L} [\hat{n}_{1}] = \Gamma (1 - \hat{n}_{1})$ and the current out of the first site is, $\hat{J}_{1,out} = i [\hat{\calH}, \hat{n}_{1}]$. Similarly, for the last site connected to the right lead, the current into the last site is, $\hat{J}_{L,in} = i [\hat{\calH}, \hat{n}_{L}]$ and the current out of the last site is $\hat{J}_{L,out} = \calD_{R} [\hat{n}_{L}] = \Gamma \hat{n}_{L}$. For all other sites, i.e., $2\leq j \leq L-1$, $\hat{J}_{j,in} - \hat{J}_{j,out} = i [\hat{\calH}, \hat{n}_{j}]$ which gives,
\begin{eqnarray}
\hat{J}_{j,in} = -i \sum_{r} \frac{J}{r^{\alpha}} \left(\hatc_{j}^{\dagger} \hatc_{j-r} - \hatc_{j-r}^{\dagger} \hatc_{j} \right), \nonumber \\ 
\hat{J}_{j,out} = -i \sum_{r} \frac{J}{r^{\alpha}} \left(\hatc_{j+r}^{\dagger} \hatc_{j} - \hatc_{j}^{\dagger} \hatc_{j+r} \right).
\end{eqnarray}
Note that the dephasing dissipator commutes with the local number operator and, therefore, does not appear in the corresponding equation of motion.

\begin{figure}
	\centering
	\includegraphics[scale=0.3]{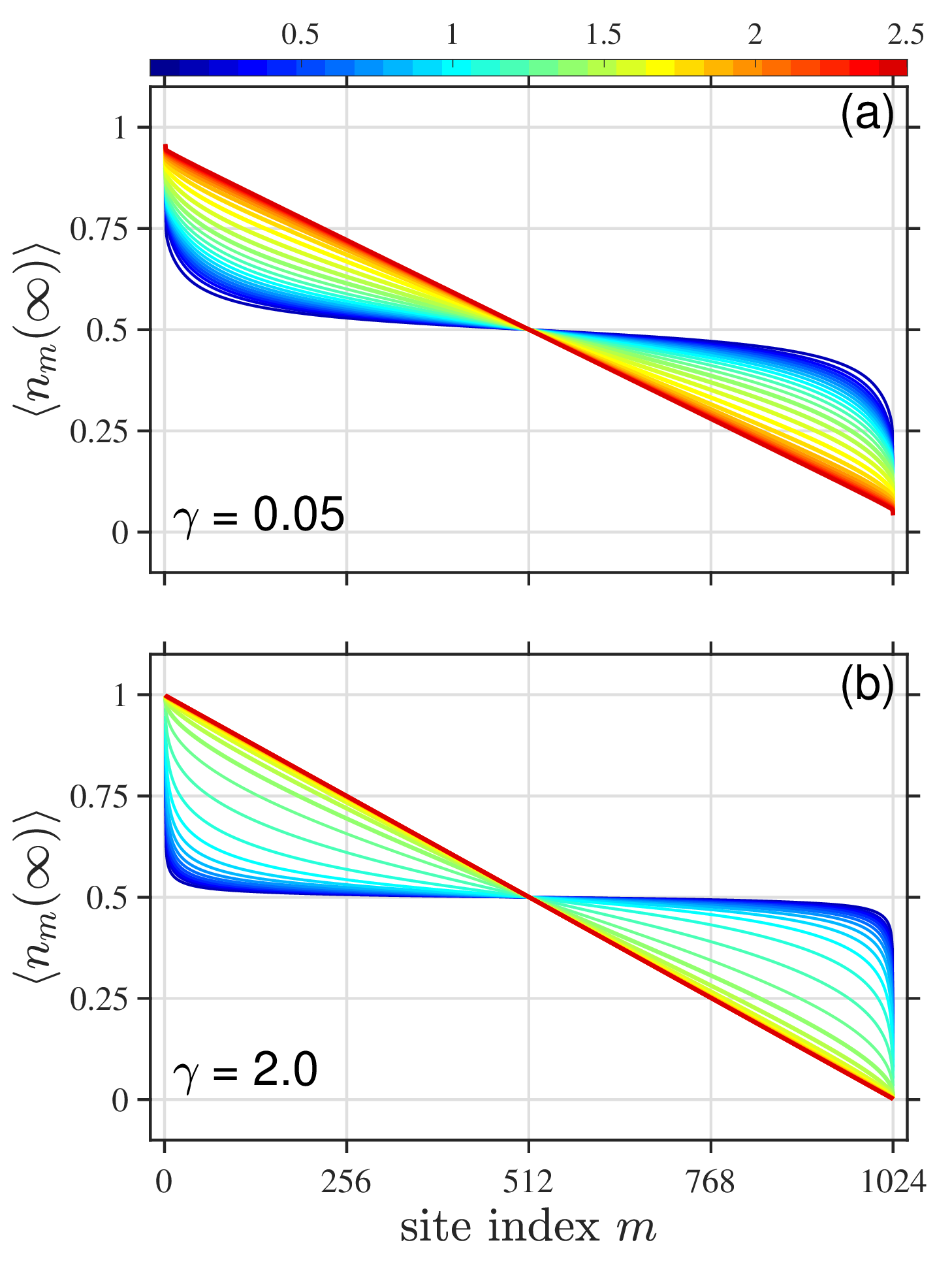}
	\caption{{Non-equilibrium} steady-state  density profile, viz., $\langle \hat{n}_{m}(\infty)\rangle$ with respect to the {lattice} site index, and {for a range of long-range hopping parameter $\alpha$}. Upper (a) and lower (b) panels correspond to {two different} dephasing strengths, $\gamma = 0.05$ and $\gamma=2.0$, respectively.} \label{fig:density}
\end{figure}

Our approach is based on analyzing the single particle correlation matrix $\mb{C}$ with matrix elements $ C_{n,m}(t) = \text{Tr} [\rho(t) \hat{c}_{n} \hat{c}_{m}^{\dagger}]$, which for the non-interacting systems can be computed more efficiently \cite{Elenewski_corr_2017, Zwolak_corr_2020, V_K_Varma_PhysRevE, Turkeshi_PhysRevB}. Moreover, it allows calculating the non-equilibrium steady state density profile and the current even for on-site dephasing cases. The equation of motion for the correlation matrix is given by,
\begin{equation}\label{eq:eq_mot_C}
\frac{\partial \mb{C}}{\partial t} = -i [\mb{H},\mb{C}] - \lbrace \mb{D},\mb{C} \rbrace + \mb{P},
\end{equation}
where $\displaystyle (\mb{D})_{m,k} = \frac{\delta_{m,k}}{2} \left(\gamma + \Gamma[\delta_{m,1} + \delta_{m,N}] \right)$, and $\displaystyle (\mb{P})_{m,k} = \delta_{m,k} \left(\gamma C_{m,n}(t) + \Gamma \delta_{m,N}\delta_{k,N} \right)$. Furthermore, \eqref{eq:eq_mot_C} can be re-written as $\displaystyle \frac{\partial \mb{C}}{\partial t} = -i \mb{H}_{\text{eff}} \mb{C} + i \mb{C} \mb{H}_{\text{eff}}^{\dagger} + \mb{P}$, where $\mb{H}_{\text{eff}}$ corresponds to the effective {single-particle} non-Hermitian Hamiltonian,
\begin{eqnarray}
\calH_{\text{eff}} &=& \hat{\calH} - i\frac{\gamma}{2} \sum_{j=1}^{L} \hat{n}_{j} -  i\frac{\Gamma}{2} \left([1-\hat{n}_{1} + \hat{n}_{L}] \right) \nn \\
& =& \sum_{i,j=1}^{L} \hatc_{i}^{\dagger}  \mb{H}^{i,j}_{\text{eff}} \hatc_{j}.
\end{eqnarray}

The spectral properties of this non-Hermitian Hamiltonian determine the stationary values of the particle density and current \cite{Elenewski_corr_2017, Zwolak_corr_2020, V_K_Varma_PhysRevE, Turkeshi_PhysRevB}. The stationary state solution corresponds to 
$\frac{\partial \mb{C}}{\partial t} = 0$ is given by, $\displaystyle \mb{C} (\infty) = \int_{0}^{\infty} d \tau e^{-i\mb{H}_{\text{eff}} \tau } \mb{P} (\infty) e^{i\mb{H}_{\text{eff}}^{\dagger} \tau }$. Then the eigendecomposition of $\mb{H}_{\text{eff}}= \sum_{p} \lambda_{p} |\phi_{R}(p)\rangle \langle \phi_{L} (p) |$ in terms of bi-orthogonal eigenvectors  $\langle\phi_{L}(p)|\phi_{R} (q) \rangle = \delta_{p,q}$, leads to the steady state values of the matrix elements of $\mb{C}$ given by,
\begin{eqnarray}\label{eq:SS_matrix_form}
C_{i,j}(\infty) &=& \Gamma\Theta_{i,j,L} + \gamma \sum_{k=1}^{L} \Theta_{i,j,k} C_{k,k}(\infty) \nonumber \\
\Theta_{i,j,k} &=& -\sum_{p,q=1}^{L} \frac{\phi_{i}^{R} (p) [\phi_{l}^{L} (p)]^{*} [\phi_{j}^{R} (q)]^{*} \phi_{l}^{L} (q)}{i(\lambda_{p} - \lambda_{q}^{*})}.
\end{eqnarray}
Using \eqref{eq:SS_matrix_form} one can obtain the NESS charge density and current using 
\begin{eqnarray}
\langle \hat{n}_{m}(\infty)\rangle &=& 1- C_{m,m}(\infty), \\ J_{\infty} = \langle \hat{J}_{m}(\infty)\rangle &=& \sum_{r} \frac{2J}{r^\alpha} \Im[C_{m,m-r}(\infty)].
\end{eqnarray}
In what follows, we will use current and resistance ($R_{\infty}= J_{\infty}^{-1}$) interchangeably in analyzing NESS transport and show our numerical findings.

\begin{figure}
	\centering
	\includegraphics[scale=0.3]{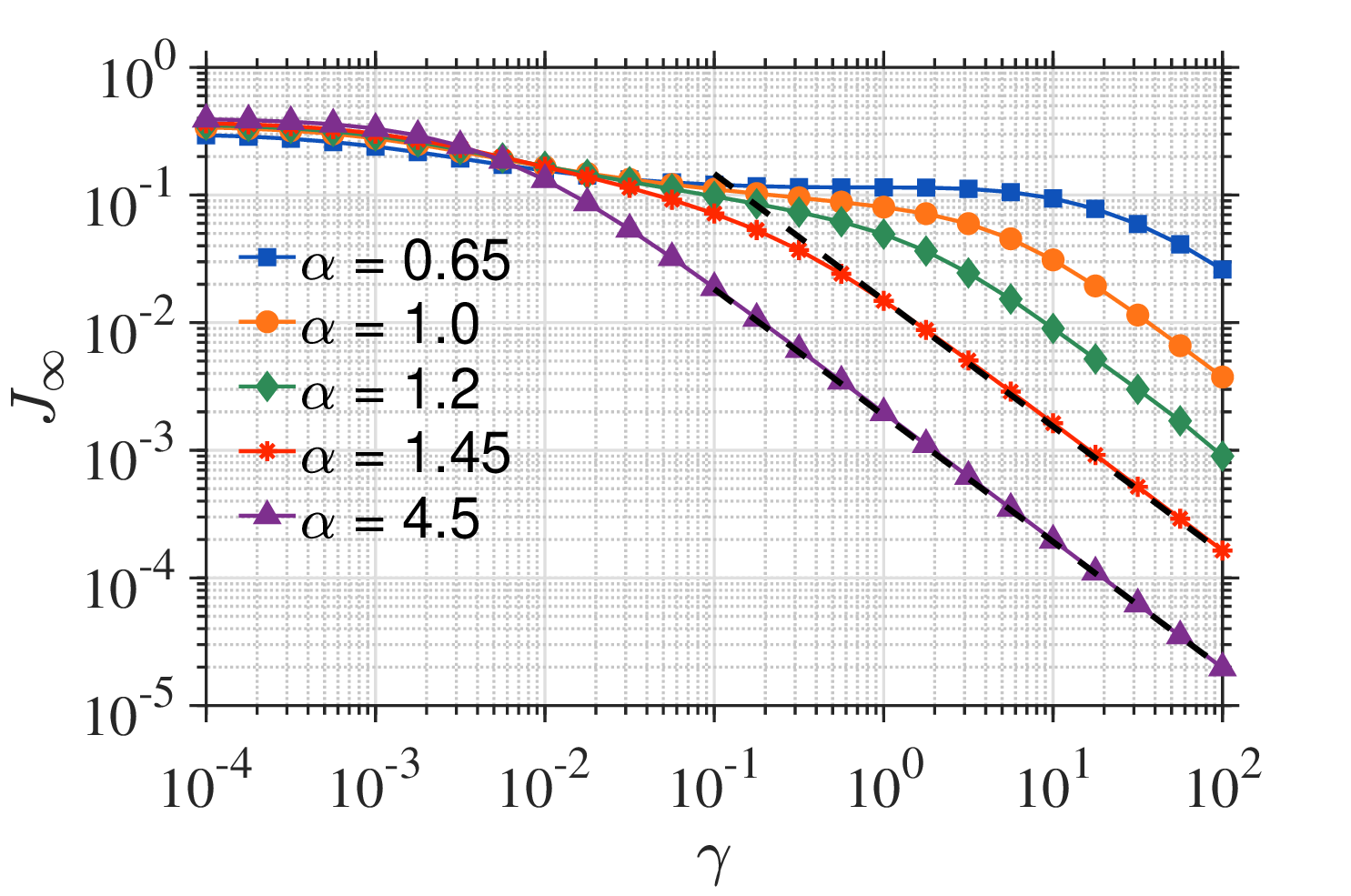}
	\caption{Non-equilibrium steady-state current $J_{\infty}$ corresponding to several values of $\alpha$ as a function of dephasing strength $\gamma$ for system size $L=1024$, in log-log scale. Black dashed lines correspond to $J_{\infty} \sim \frac{1}{\gamma}$ scaling. 
 }
	\label{fig:J_vs_gamma}
\end{figure} 
\begin{figure*}[!t]
	\centering
	\includegraphics[scale=0.4]{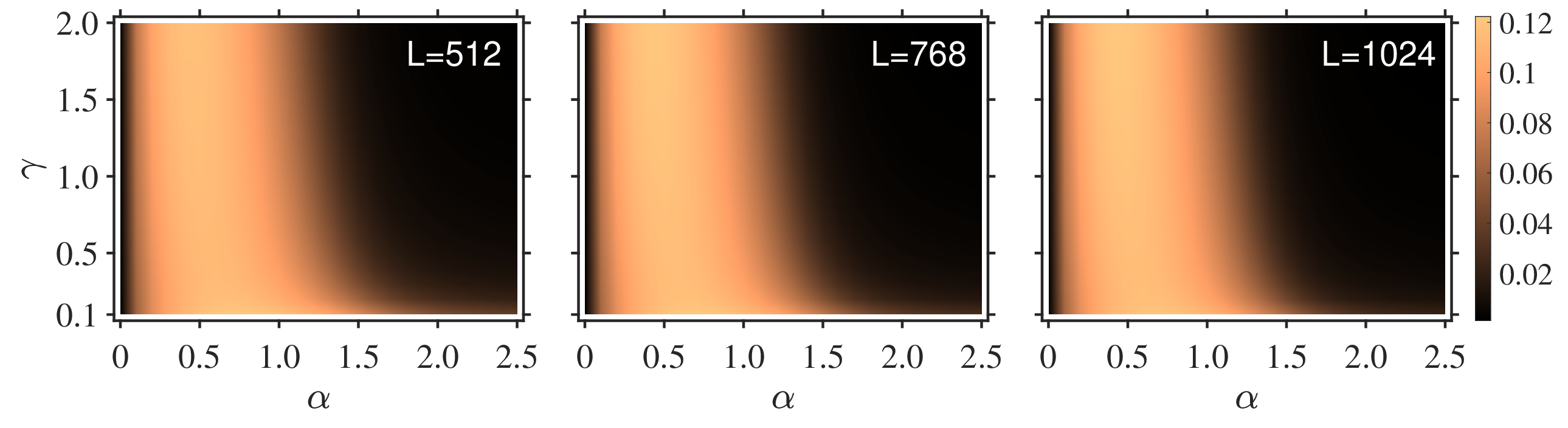}
	\caption{The heat map of current $J_{\infty} = \langle \hat{J}_{L,out}\rangle$. x- and y-axis plot long-range hopping exponent $\alpha$ and dephasing $\gamma$, and the color-bar represents the magnitude of  $J_{\infty}$ for three different length of the lattice.
	} \label{fig:J_vs_alpha_vs_dephase}
\end{figure*}
\section{Results} \label{Sec:results}
We focus on the NESS density profile throughout the lattice chain and the transport current from the right lead, viz., $J_{\infty} = \Gamma \langle \hat{n}_{L}(\infty)\rangle$. We study the system size scaling of the NESS current to characterize the various transport regimes. 
We fix the hopping amplitude $J=1$. The dephasing strength $\gamma$ and particle injection-extraction rate $\Gamma$ are taken in the unit of $J$. In all the following analysis, we consider the system-lead coupling to be $\Gamma = 1$. It is a rather benign parameter compared to $\gamma$ and $\alpha$ within our infinite bias set-up as it only determines the overall magnitude of the current.

\textit{{NESS density profile and current}}: We first study the steady state density profile of the system and plot $\langle \hat{n}_{m}(\infty)\rangle$ as a function of the site index `$m$' for a range of long-range parameter $\alpha$ (shown in the color bar) in Fig.~\ref{fig:density}. We consider the dephasing strengths $\gamma = 0.05$ [upper panel (a)] and $\gamma = 2.0$ [lower panel (b)], respectively. For $\alpha \neq 0$, and small dephasing strength $\gamma = 0.05$, the density profile shows a trend towards a linear profile for larger $\alpha$-values starting from a non-linear profile at smaller $\alpha$-values. Although, $\langle \hat{n}_{m}(\infty)\rangle$ shows the same trend towards a linear profile for larger dephasing, e.g., $\gamma = 2.0$ in Fig.~\ref{fig:density}, interestingly, $\langle \hat{n}_{m}(\infty)\rangle$ is more flattened in the bulk of the chain compared to the level of flatness seen for $\gamma = 0.05$, thereby suggesting an enhanced transport due to the dephasing. This is opposite to the expectation, i.e., in the absence of dephasing, a chain with nearest-neighbor hopping exhibits a flat density profile in bulk with $\langle n_m(\infty)\rangle\approx 0.5$, a hallmark of ballistic transport and, for stronger dephasing strength, the density profile scales as $\langle n_m (\infty)\rangle \propto 1/L$ suggesting a suppression in transport (diffusive transport). Such contrasting features hint towards {the emergence of an interesting} anomalous transport regime in the presence of long-range hopping.

To explore the regime of unusual transport, we study the variation of the NESS current $J_{\infty}$ for different values of dephasing $\gamma$ and long-range parameter $\alpha$ for a fixed system size $L =1024$. Fig. \ref{fig:J_vs_gamma} plots $J_{\infty}$ as a function of dephasing strength $\gamma$. We observe that the current gets suppressed with increased dephasing for all the values of $\alpha$. This is clearly seen in Fig. \ref{fig:J_vs_gamma} (black dashed lines) for $\alpha = 1.5$ and higher, where current decreases monotonically with $\gamma$, i.e., $J_{\infty} \propto \frac{1}{\gamma}$. {For the short-range hopping model, in the absence of dephasing, the transport is always ballistic. Finite dephasing is detrimental to this ballistic transport and plays a role in in-elastic scattering \cite{Turkeshi_PhysRevB}.} On the other side, for $\alpha \ll 1$, current initially decreases with increasing dephasing and, interestingly, settles down to a plateau regime for a considerable range $10^{-2} \lesssim \gamma \lesssim 10^{1}$ of dephasing. With increasing $\alpha$, the range of $\gamma$ over which this interesting plateau regime appears starts to shrink and eventually disappears for $\alpha \geq 1.5$. The plateau regime emerges due to the intricate interplay of long-range hopping and dephasing. 
In other words, due to long-range hopping, particles can now evade the in-elastic scattering induced by the dephasing more easily and, thereby, can possibly deviate from standard diffusive transport.

In the following, we elaborate further on the appearance of the plateau and restrict ourselves to the dephasing strength $0.1<\gamma < 2.0$, for which the $J_{\infty}$ plateaued in Fig. \ref{fig:J_vs_gamma}. We plot a heat map of the $J_{\infty}$ as a function of long-range parameter $\alpha$ and dephasing $\gamma$ in Fig. \ref{fig:J_vs_alpha_vs_dephase}, for system sizes $L = 512,~ 768$, and $1024$, respectively. Clearly, comparing the plots corresponding to all these chain lengths, we find that apart from the magnitude of $J_{\infty}$, other features do not distinctively differ. This indicates no significant finite size effect in $J_{\infty}$ with the system size in the physical picture to follow. Interestingly, the heat map of the NESS current on the $\alpha - \gamma$ plane closely resembles the phase diagram of entanglement entropy, which characterizes the phenomenon of measurement-induced phase transition \cite{minato2022fate}, and thus hints at a possible interesting connection between the MIPT and quantum transport. 

Firstly, we observe that NESS current becomes almost negligible for very small values of $\alpha \rightarrow 0$. In this limit, the system manifests all-to-all coupling that hinders transport and incites cooperative shielding, akin to Anderson localization \cite{celardo2016shielding}. Intriguingly, this cooperative shielding is robust against the dephasing and system size $L$, as evident from Fig. \ref{fig:J_vs_alpha_vs_dephase}. Then, for a finite $\alpha$, we note a surge in NESS current, peaking near $\alpha \approx 0.6$ for values of $\gamma>0.1$ and remains almost independent of $\gamma$. 
Recall that this is reminiscent of the plateau observed in Fig. \ref{fig:J_vs_gamma} for larger $\gamma$ values for $\alpha = 0.65$. On further increasing the value of $\alpha$, the plateau regime shrinks and eventually disappears for $ \alpha>1.5 $ irrespective of $\gamma$, as seen from Fig. \ref{fig:J_vs_alpha_vs_dephase}.%

\begin{figure}[t]
	\centering
	\includegraphics[scale=0.3]{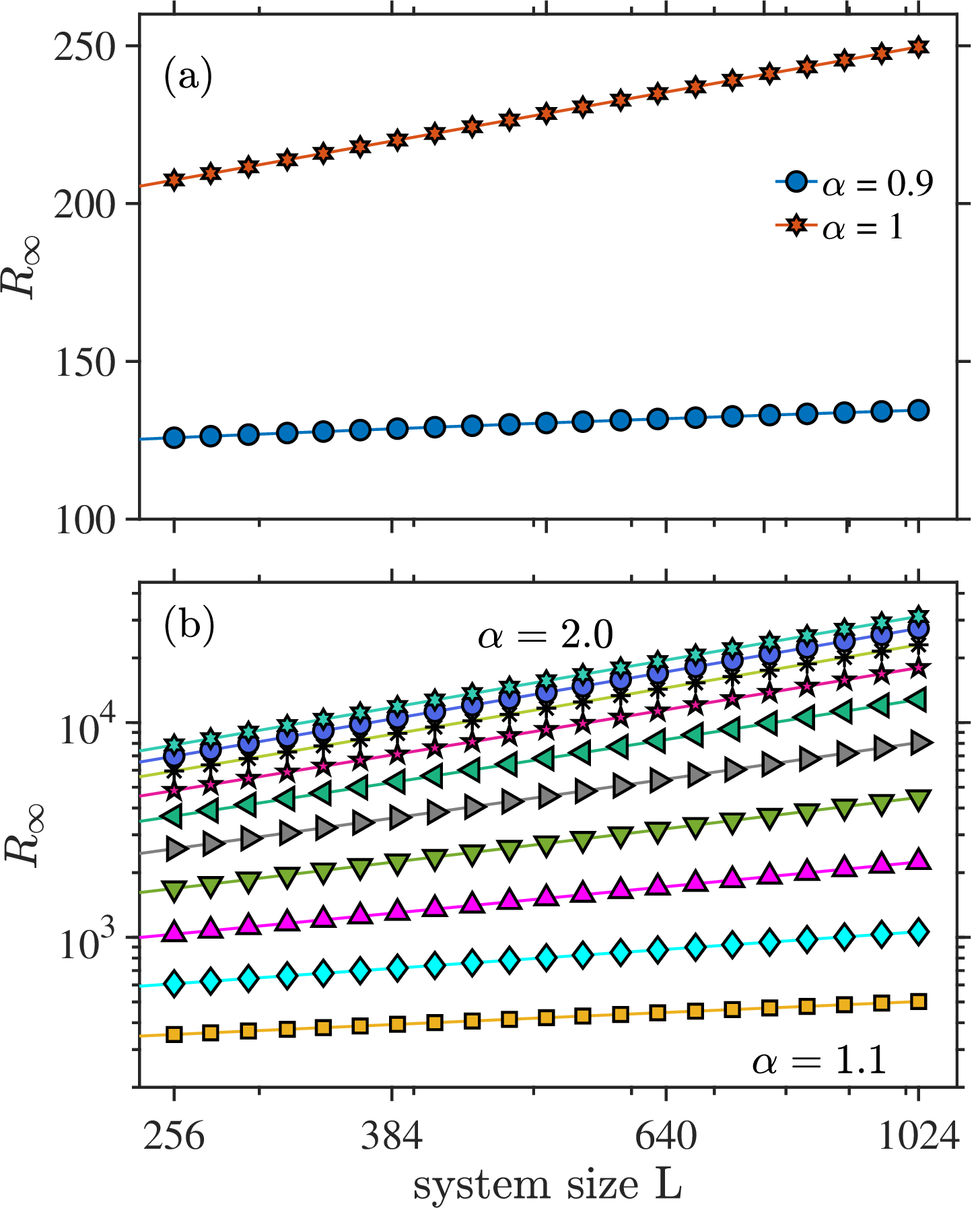}
	\caption{NESS resistance $R_{\infty} = 1/ J_{\infty}$ as a function of system size $L$, (a) on a log-linear scale for $\alpha = 0.9$ and $1.0$, (b) on a log-log scale from $\alpha = 1.1$ to $2.0$ (bottom to top) in an increment of $0.1$.
 }
\label{fig:R_vs_L_gamma}
\end{figure}
 
\textit{Nature of the NESS transport}: 
To understand the nature of the underlying NESS transport, we examine the system size scaling of the NESS resistance, denoted as $R_{\infty} = 1/J_{\infty}$, for various values of $\alpha$. Note that the scaling results presented below are almost insensitive for $\gamma > 0.1$. We, therefore, present results for $\gamma = 10$. Depending on the system size scaling of $R_{\infty}$ different transport regimes are classified, namely, for ballistic transport $R_{\infty} \sim L^{0}$ \cite{dhar2006heat}, for diffusive or normal transport $R_{\infty} \sim L$ \cite{Turkeshi_PhysRevB}. Anything away from these scaling relations is often categorized as anomalous transport, for example, $R_{\infty} \sim L^{\nu}$ with $\nu >1$ it is sub-diffusive \cite{Purkayastha_subdiffusive_2021} whereas for $\nu < 1$ it is super-diffusive.

In Fig.~\ref{fig:R_vs_L_gamma}, we clearly demonstrate the system size scaling of $R_\infty$ for different values of $\alpha$ to analyze the regimes of transport. In Fig.~\ref{fig:R_vs_L_gamma}(a) we first concentrate on $\alpha \leq 1$. We plot $R_\infty$ as a function of system size $L$ for $\alpha = 0.9$ and $\alpha = 1.0$. Interestingly, we observe logarithmic system size scaling, $R_{\infty} \sim \log(L)$ as clear from the straight-line fitting of the numerical data on a log-linear scale. This clearly demonstrates an anomalous super-diffusive transport regime for $\alpha \leq 1$. Next, in Fig.~\ref{fig:R_vs_L_gamma}(b) we focus on the case of $\alpha >1$. We plot $R_{\infty}$ as a function of $L$ for $1.0 < \alpha \leq 2$. Remarkably, in this regime of $\alpha$ we observe a power-law dependence: $R_{\infty} \sim L^{\nu}$ with an $\alpha$-dependent exponent $\nu$. This is evident from the plot of $R_{\infty}$ on a log-log scale.
Therefore, it is clear from Fig.~\ref{fig:R_vs_L_gamma}(a) and (b) that as $\alpha$ increases beyond the value 1, the scaling of $R_{\infty}$ changes from a logarithmic to a power-law behavior in system size. 

\begin{figure}[b]
	\centering
    \includegraphics[scale=0.3]{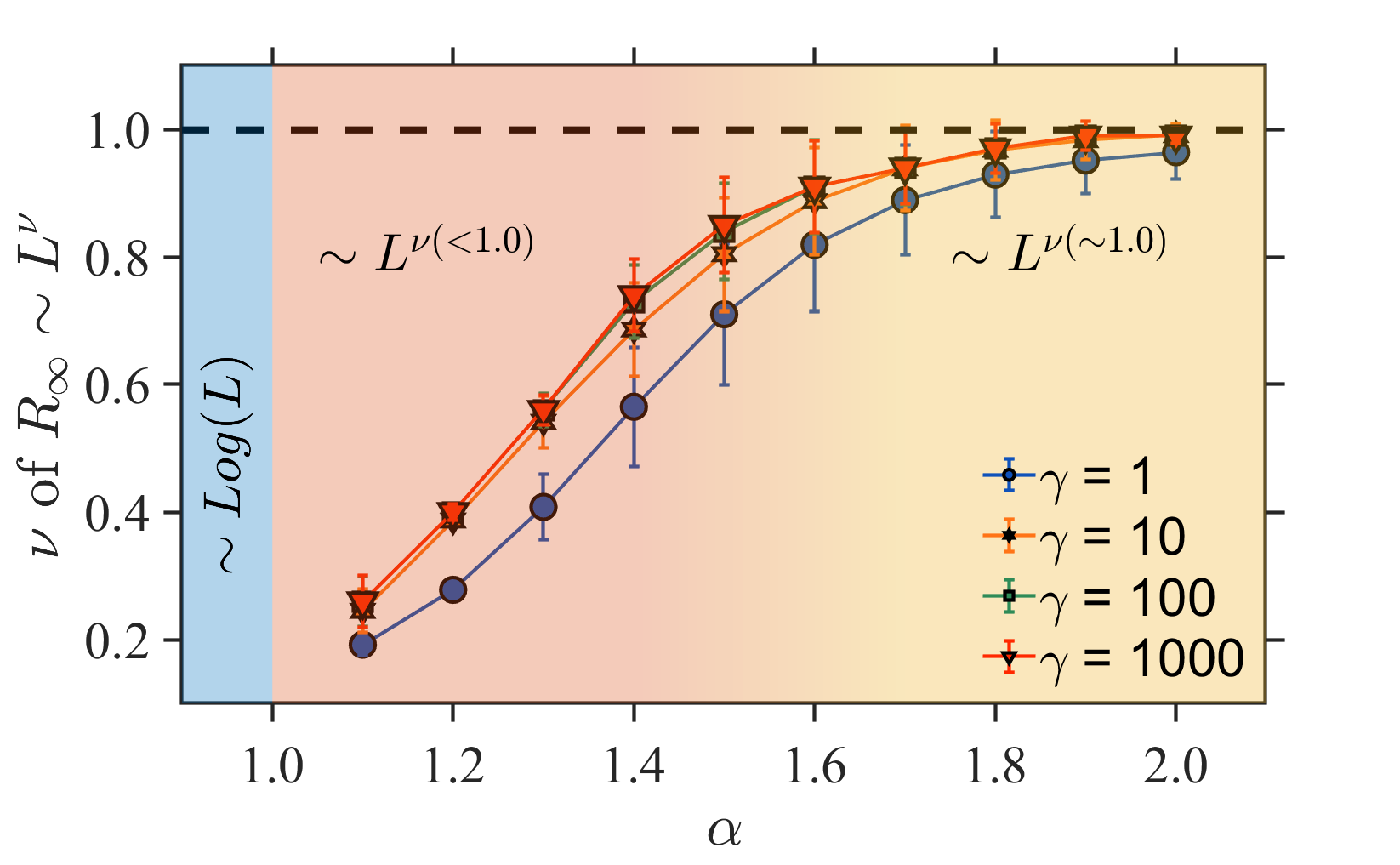}
	\caption{Plot of the system size scaling exponent $\nu$ corresponding to the fit $R_{\infty} \sim L^{\nu}$ as a function of $\alpha$ for $\alpha > 1.0$ and different values of dephasing strength $\gamma$. The error bars are associated with the fitting of $R_{\infty} \sim L^{\nu}$.}
	\label{fig:Slope_R_vs_L_gamma}
\end{figure}

To further elaborate on the nature of the transport for $\alpha > 1$, we plot the dependence of the exponent $\nu$ with $\alpha$  in Fig. \ref{fig:Slope_R_vs_L_gamma}. Our numerical calculations show that $\nu$ increases with increasing $\alpha$ before saturating beyond $\alpha \gtrsim 1.6$. It is worth mentioning that the super-diffusive transport for $0.0 < \alpha \lesssim 1.6$ is extremely robust against the dephasing strength. However, a finite amount of dephasing along with the long-range hopping is necessary to observe the emergence of these intriguing transport regimes. Remarkably, for $\alpha \gtrsim 1.6$, a diffusive transport regime sets in independent of the value of dephasing. However, we show below that the transition from super-diffusive to diffusive transport in the thermodynamic limit happens exactly at $\alpha = 1.5$. We believe this difference in the value of $\alpha$ appears due to the finite size effect. 

To substantiate the above numerical observation that a diffusive transport emerges only for $\alpha>1.5$, we show that the behavior of the current operator norm (in the absence of the dephasing), analogously to the Hamiltonian operator norm discussed in Ref. \cite{minato2022fate}, can provide more insightful information. We start by writing down the operator for current into the site $L$ as 
\begin{eqnarray}\label{eq:current_op}
\hat{J}_{L,in} &=& -i \sum_{r=1}^{L-1} \frac{J}{r^{\alpha}} \left(\hatc_{L-r}^{\dagger} \hatc_{L} - \hatc_{L}^{\dagger} \hatc_{L-r} \right).
\end{eqnarray}
The corresponding operator norm $||J_L||=\text{Tr}\left( \sqrt{\hat{J}_{L}^\dagger \hat{J}_{L}}\right)$ exhibit the following system-size dependence (see Appendix \ref{app:op_norm} for the details of the calculation),
\begin{eqnarray} \label{eq:scaling}  ||J_{L}|| &\sim& \frac{L^{3/2 - \alpha}}{\sqrt{(2\alpha -1)}(3/2 - \alpha)},  ~ \text{for } \alpha < 1.5, \nn \\
    &\sim& \text{constant independent of $L$ for } \alpha >1.5.
\end{eqnarray}
This suggests that the current operator norm, representing the maximum rate of coherent particle transport, remains constant and independent of $L$ for $\alpha>1.5$. Therefore, for a finite system size, a sufficiently strong dephasing strength can suppress coherent transport and lead to diffusive behavior. In fact, the scaling behavior of $||J_{L}||$ mentioned above implies that, for $\alpha > 1.5$ in the thermodynamic limit, any non-zero dephasing can induce diffusion similar to the short-range hopping model \cite{Znidaric_2010_J_Stat_Mech}.
\paragraph*{}
On the other hand, the current operator norm diverges when $\alpha<1.5$, implying that, in the thermodynamic limit, any amount of dephasing strength is insufficient to yield diffusive (or slower than diffusive, e.g., sub-diffusive) transport. Furthermore, the divergent system size scaling also suggests that, in the presence of finite (and large) dephasing, transport cannot be ballistic either in the thermodynamic limit. This leaves us with the only possibility of super-diffusive transport in the thermodynamic limit, even in the presence of finite (and large) dephasing. Therefore, the behavior of the operator norm in the absence of dephasing suggests that $\alpha=1.5$ holds a special significance, aiding our understanding of the transition from super-diffusive to diffusive transport in the presence of dephasing.

\section{Summary and Discussion} In summary, using the Markovian Lindblad quantum master equation, we examined the NESS transport properties of non-interacting fermions on a one-dimensional lattice with \( \frac{1}{r^\alpha} \)-long-range hopping, subjected to on-site dephasing. A non-trivial interplay between the long-range hopping and the dephasing strength leads to the emergence of distinctive regimes of NESS transport as one tunes the long-range exponent $\alpha$. Specifically, we demonstrate that for \( \alpha > 1.5 \), the transport is diffusive/normal, whereas, for $\alpha < 1.5$, the transport is super-diffusive. Furthermore, we observe two different system-size scaling within the super-diffusive regime; namely, for $1<\alpha<1.5$, we find a power law system size dependence with an \( \alpha \)-dependent exponent \( \nu < 1 \) and for \( \alpha \leq 1 \), we observe inverse-logarithmic system size scaling for the current. Remarkably, the super-diffusive regime is robust against the dephasing strength. We further provide analytical insights to support our numerical findings by analyzing the current operator norm. 

It is worth pointing out that the heat map of the NESS current on the $\alpha - \gamma$ plane, as presented in Fig.~\ref{fig:J_vs_alpha_vs_dephase}, bears a close resemblance to the phase diagram of entanglement measures related to the MIPT\cite{minato2022fate}. The NESS transport features we delved into stem from the steady-state solution of the Lindblad quantum master equation, whereas, in the context of MIPT, the transition of entanglement entropy from an area law phase to a volume law phase emerges through a specific unraveling of the Lindblad master equation, namely the quantum state diffusion. For the free Fermion case in one dimension (without the boundary drive), MIPT appears at the long-range exponent $\alpha = 1.5$~\cite{minato2022fate}. Intriguingly, in our setup, a transition from super-diffusive to diffusive regime also appears at the same long-range exponent value, i.e., $\alpha = 1.5$. Given that number-conserving local dephasing in our setup acts like a Zeno-type local number density measurement, our results hint at a connection between the MIPT and the transition from anomalous to normal observed in the NESS current. 

However, our identification of transport as a possible signature of the underlying MIPT is not caveat-free. Indeed, the absence of a direct link between mutual information and particle current renders such an identification non-trivial, positing an intriguing avenue for future research. Relevantly, prior numerical analyses have pointed out interesting connections between NESS transport current and entanglement measures, like mutual information and concurrence, particularly in quantum dot set-ups~\cite{sharma2015landauer,sable2018landauer,dey2019quantum}. Establishing a more rigorous link could pave the way for discerning MIPT in experiments via current measurements.

 \appendix 
 \section{Operator norm and explanation of super-diffusive to diffusive crossover} \label{app:op_norm} 
\begin{figure}
    \centering
    \includegraphics[scale=0.26]{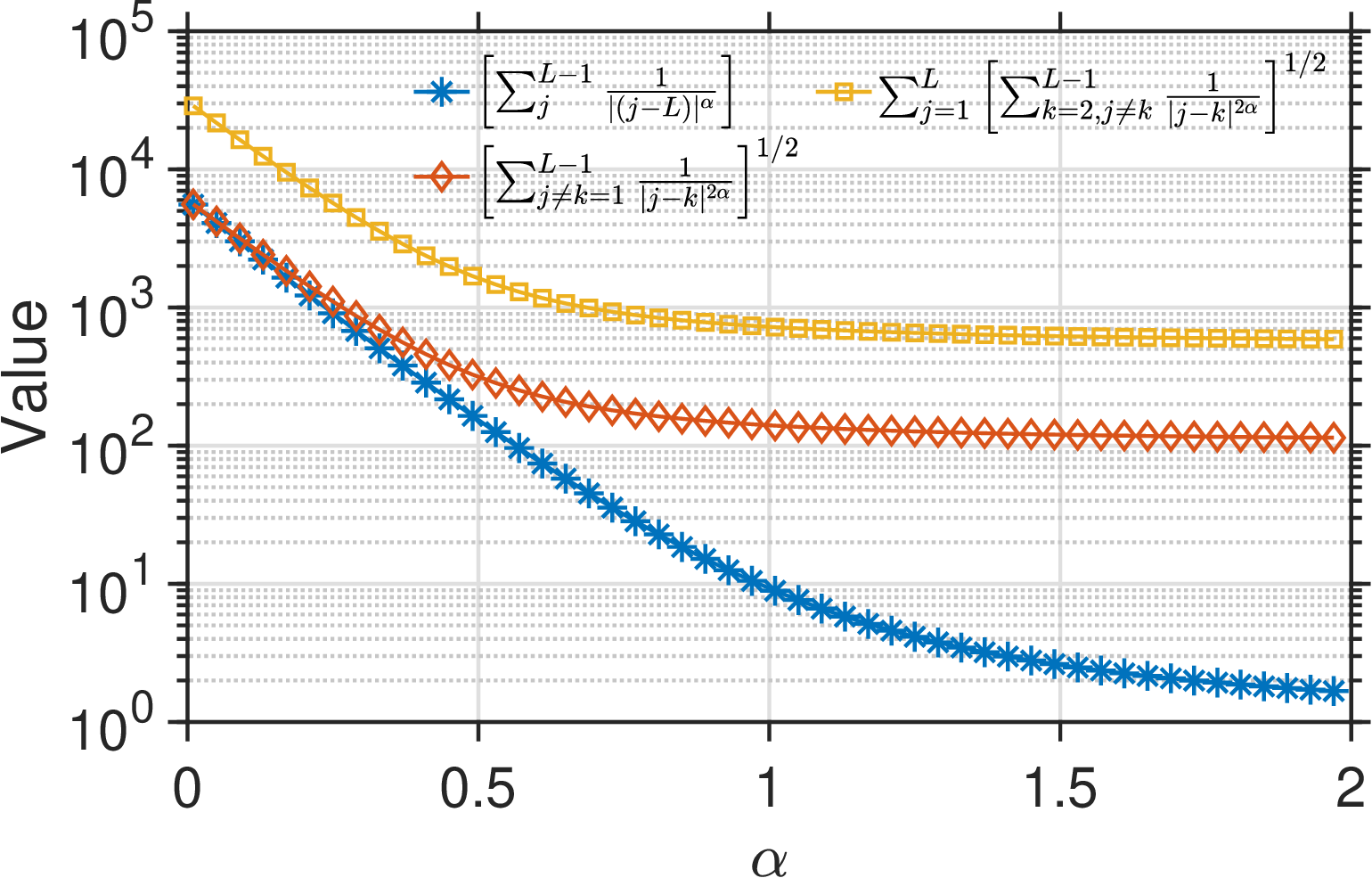}
    \caption{Numerical value of the upper bound of the operator norm verifying the inequalities in  \eqref{eq:op_norm_ineq}, for $L=6000$. 
    }
    \label{fig:operator_norm}
\end{figure}

Here we would like to prove the scaling of $||J_L||$ corresponding to \eqref{eq:scaling}. We start by writing down the operator for current into the site $L$ as 
\begin{eqnarray}\label{eq:current_op}
\hat{J}_{L} &=& -i \sum_{r=1}^{L-1} \frac{1}{r^{\alpha}} \left(\hatc_{L-r}^{\dagger} \hatc_{L} - \hatc_{L}^{\dagger} \hatc_{L-r} \right), \nn \\ &=& -i  \sum_{j=1}^{L-1} \frac{1}{(j-L)^{\alpha}} \left(\hatc_{j}^{\dagger} \hatc_{L} - \hatc_{L}^{\dagger} \hatc_{j} \right) \nn \\ &=& -i  \sum_{j=1}^{L-1} \frac{1}{(j-L)^{\alpha}} \hat{O}_{j}.
\end{eqnarray} 
We can then calculate the upper bound of the operator norm with respect to the trace norm $||J_L|| = \text{Tr} \left(\sqrt{\hat{J}_{L}^{\dagger} \hat{J}_{L}}\right)$ which is given by, \begin{eqnarray} \label{eq:op_norm_ineq_0}
    ||J_L||  &\leq& \sum_{j=1}^{L-1} \frac{1}{|j-L|^{\alpha}} ||\hat{O}_{j}||
    \leq \sum_{j=1}^{L-1} \frac{1}{|j-L|^{\alpha}} \nn \\
    &=& \left[\sum_{j,k=1}^{L-1}\frac{1}{|(j-L)(k-L)|^\alpha }\right]^{1/2},
\end{eqnarray} 
where we have used the triangle inequality for the operator norm, $\displaystyle || \sum_{j} \hat{A}_{j} || \leq \sum_{j} || \hat{A}_{j} ||$, where $\hat{A}_{j}$ is an operator defined on a normed Hilbert space, in obtaining the first inequality in \eqref{eq:op_norm_ineq_0}, and also used the upper bound of the operator norm $||\hat{O}_{j}|| \leq 1$. 
Next, we can show that the following inequalities hold,
\begin{eqnarray}\label{eq:op_norm_ineq}
    ||J_L|| &\leq& \sqrt{\sum_{j,k=1}^{L-1}\frac{1}{ |(j-L)(k-L)|^\alpha }} \leq \sqrt{\sum_{j=1}^{L-1} \sum_{\substack{k=1 \\ j\neq k}}^{L-1}\frac{1}{|j-k|^{2\alpha}}}. \nn \\  &\leq& \sum_{j=1}^{L-1} \left[\sum_{\substack{k=1 \\ k \neq j}}^{L-1}\frac{1}{|j-k|^{2\alpha}}\right]^{1/2},
\end{eqnarray}
We further confirm this by plotting the various summations of Eq.\eqref{eq:op_norm_ineq} in Fig.~\ref{fig:operator_norm}. It can be seen that the inequality indeed holds for every value of $\alpha$ and differs only by a constant factor that is independent of the system size.

Converting the discrete summations of \eqref{eq:op_norm_ineq} to integral we find in the large $L$ limit, \begin{eqnarray}\label{eq:scaling_1} & &||J_{L}|| \leq \int_{x=1}^{L}  dx \left[\int_{y=2}^{L-1}\frac{dy}{|x-y|^{2\alpha}}\right]^{1/2} \nn \\ &=&  \left[\int_{y=2}^{L-1}\frac{dy}{|1-y|^{2\alpha}}\right]^{1/2} +  \int_{x=2}^{L} dx \left[\int_{y=1}^{L-1}\frac{ dy}{|x-y|^{2\alpha}}\right]^{1/2} \nn \\ &\approx& \sqrt{\frac{(1-L^{1-2\alpha})}{(2\alpha -1 )}} + \frac{L^{3/2 - \alpha}}{\sqrt{(2\alpha -1)}(3/2 - \alpha)},  \end{eqnarray} where we have identified in the continuum limit $i\rightarrow x$ and $j\rightarrow y$, and implicitly assumed $\alpha > 0.5$. 

 For $\alpha \geq 1.5~ (=3/2+\epsilon)$ (with $\epsilon \geq 0$ is just a parameter) the first term gives, $\sim \frac{1}{\sqrt{2\alpha - 1}}$ and the second term gives, $-\frac{1}{L^{\epsilon}\sqrt{(2\alpha -1)}\epsilon}$, making the second term vanishing faster and leaving the dominant scaling to be a constant independent of the chain-length $L$. For $\alpha < 1.5$ in equation \eqref{eq:scaling_1}, the dominant scaling comes from, $\displaystyle \sim \frac{L^{3/2 - \alpha}}{\sqrt{(2\alpha -1)}(3/2 - \alpha)}$.

 Therefore, we arrive at the following important leading order behavior of the current operator norm, \begin{eqnarray} ||J_{L}|| &\sim& \frac{L^{3/2 - \alpha}}{\sqrt{(2\alpha -1)}(3/2 - \alpha)}  ~ (0 \leq \alpha < 1.5 ) \nn \\
    &\sim& L~\text{independent, for } \alpha >1.5.
\end{eqnarray} This operator norm scaling signifies that for $\alpha > 1.5$ the current operator norm is upper bounded by a $L-$independent constant whereas, for $\alpha < 1.5$, it is upper bounded by a power-law scaling with the system size, a representative of a divergent behavior. 

\section*{Acknowledgment}
SS acknowledges G. Wójtowicz, M. Zwolak, M. M. Rams, M. Dalmonte, and Y. Dubi for useful comments and suggestions, and funding from the National Science Center, Poland, under project 2020/38/E/ST3/00150. BKA acknowledges the MATRICS grant MTR/2020/000472 from SERB, Government of India, and the Shastri Indo-Canadian Institute for providing financial support for this research work in the form of a Shastri Institutional Collaborative
Research Grant (SICRG). BKA. would also like to acknowledge funding from the National Mission on Interdisciplinary  Cyber-Physical  Systems (NM-ICPS)  of the Department of Science and Technology,  Govt.~of  India through the I-HUB  Quantum  Technology  Foundation, Pune, India. BKA and SS would like to thank the International Centre for Theoretical Sciences (ICTS) for organizing the program - Periodically and quasi-periodically driven complex systems (code: ICTS/pdcs2023/6) where certain interesting discussions related to this project took place. 

\bibliography{main}
\end{document}